\documentclass[aps,twocolumn,floatfix,superscriptaddress,letterpaper,longbibliography]{revtex4-2}
\usepackage{graphicx}
\usepackage{bm}
\usepackage{amsmath,amsfonts}
\usepackage{amsbsy}
\usepackage[colorlinks=true,linkcolor=blue,urlcolor=blue,citecolor=blue,breaklinks=true]{hyperref}
\usepackage[utf8]{inputenc}
\DeclareUnicodeCharacter{03BC}{\mbox{$\mu$}}
\DeclareUnicodeCharacter{2009}{ }

\begin{document}

\title{Relaxation of the degenerate one-dimensional Fermi gas}

\author{K. A. Matveev}

\affiliation{Materials Science Division, Argonne National Laboratory,
  Argonne, Illinois 60439, USA}

\author{Zoran Ristivojevic}

\affiliation{Laboratoire de Physique Th\'{e}orique, Universit\'{e} de
  Toulouse, CNRS, UPS, 31062 Toulouse, France}

\date{June 18, 2020}

\begin{abstract}
  We study how a system of one-dimensional spin-$\frac12$ fermions at
  temperatures well below the Fermi energy approaches thermal
  equilibrium.  The interactions between fermions are assumed to be
  weak and are accounted for within the perturbation theory.  In the
  absence of an external magnetic field, spin degeneracy strongly
  affects relaxation of the Fermi gas.  For sufficiently short-range
  interactions, the rate of relaxation scales linearly with
  temperature.  Focusing on the case of the system near equilibrium,
  we linearize the collision integral and find exact solution of the
  resulting relaxation problem.  We discuss the application of our
  results to the evaluation of the transport coefficients of the
  one-dimensional Fermi gas.
\end{abstract}
\maketitle

\section{Introduction}
\label{sec:introduction}

Relaxation of conventional Fermi liquids is well understood
\cite{lifshitz_statistical_1980}.  It is dominated by two-particle
collisions of the elementary excitations of the liquid.  At low
temperature $T$ the number of states available for scattering is
small, resulting in a small relaxation rate $\tau^{-1}\propto T^2$.
The fact that $\tau^{-1}$ is small compared with the typical energy
$T$ of the excitation is at the foundation of the Fermi liquid theory
\cite{lifshitz_statistical_1980}.  It is important to keep in mind
that the above result applies only to systems of fermions in two or
more spatial dimensions.

Relaxation proceeds very differently in one dimension
\cite{imambekov_one-dimensional_2012}.  Most importantly, the
scattering processes involving only two fermions do not lead to
relaxation, and thus the dominant processes involve three particles.
The relaxation rate for spin-polarized one-dimensional fermions scales
as $\tau^{-1}\propto T^7$ \cite{imambekov_one-dimensional_2012,
  arzamasovs_kinetics_2014, protopopov_relaxation_2014,
  matveev_thermal_2019}.  Such a weak relaxation at $T\to0$ is due to
the small density of states for three-particle scattering and a strong
suppression of the scattering amplitude for spin-polarized fermions,
which is a manifestation of the Pauli principle
\cite{matveev_decay_2013}.

\begin{figure}[b]
\includegraphics[width=.48\textwidth]{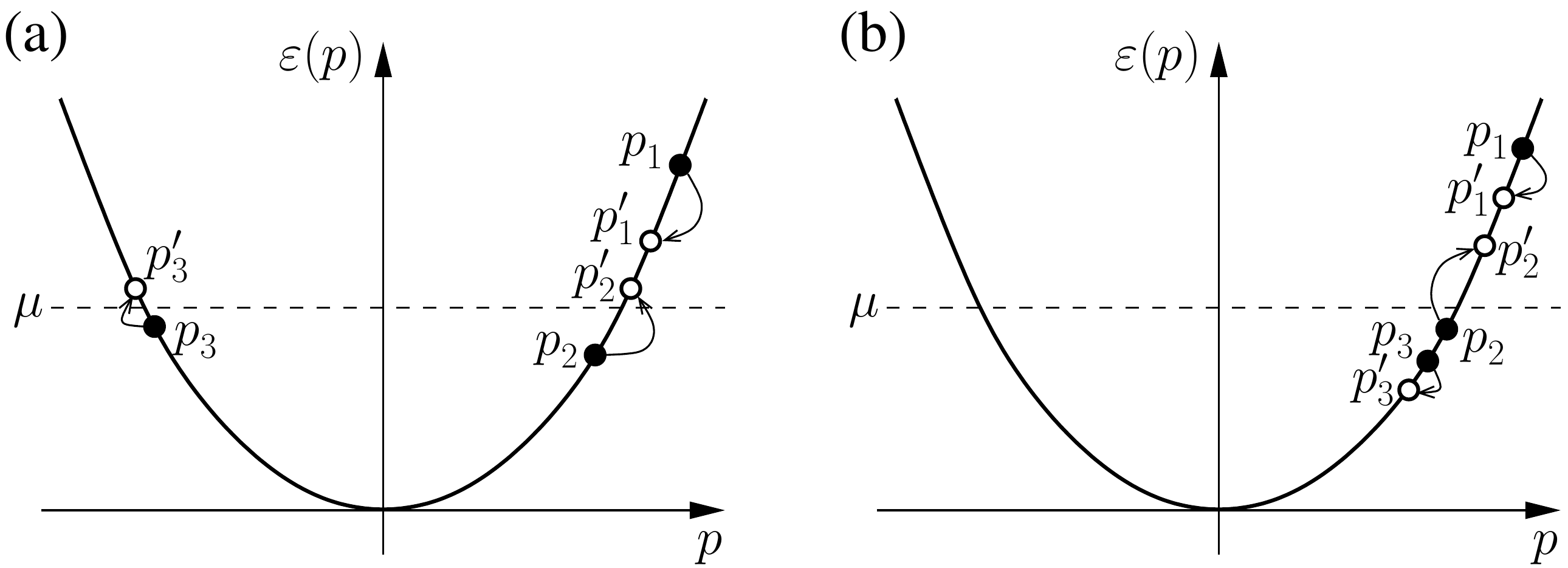}
\caption{At low temperature $T\ll\mu$ the dominant scattering
  processes involve three particles with energies within $T$ from the
  chemical potential $\mu$. (a) A typical three-particle scattering
  process with two fermions near one Fermi point and one near the
  other Fermi point.  (b) A scattering process with all three
  fermions near the same Fermi point.}
\label{fig:processes}
\end{figure}

The goal of this paper is to explore relaxation of the one-dimensional
Fermi gas in the absence of magnetic field, when the system is fully
spin-degenerate.  We will consider the low-temperature regime
$T\ll \mu$, where $\mu$ is the chemical potential.  At these low
temperatures the dominant scattering processes involve three particles
with energies near $\mu$, see Fig.~\ref{fig:processes}.  The processes
illustrated in Fig.~\ref{fig:processes}(a) involve two fermions near
one Fermi point and the third fermion near the other one.  They give
rise to decay of quasiparticles both at finite temperature and at
$T=0$.  The decay rate of a quasiparticle with energy of order $T$ due
to scattering processes of this type was evaluated in
Ref.~\cite{karzig_energy_2010}.  The result, $\tau^{-1}\propto T$, is
much greater than the decay rate $\tau^{-1}\propto T^7$
\cite{imambekov_one-dimensional_2012, arzamasovs_kinetics_2014,
  protopopov_relaxation_2014, matveev_thermal_2019} for spin-polarized
fermions, because the scattering amplitude, instead of being
suppressed due to the Pauli principle, diverges at small momentum
transfer as $|p_1-p_1'|^{-1}$.  The processes shown in
Fig.~\ref{fig:processes}(b) involve three particles near the same
Fermi point and are not allowed at zero temperature.  To our
knowledge, their effect on the decay of quasiparticles in the
spin-degenerate Fermi gas has not been considered before.  We will
show that their contribution is small compared with that of the
processes in Fig.~\ref{fig:processes}(a) only for interactions that
fall off sufficiently fast with the distance between particles.

Focusing on the latter case, we consider the relaxation of the Fermi
gas to equilibrium.  When the distribution function is close to the
equilibrium form, we are able to find a complete solution of the
relaxation problem by diagonalizing exactly the linearized collision
integral corresponding to the processes of Fig.~\ref{fig:processes}(a)
at small temperature.  This solution enables one to obtain the time
evolution of any non-equilibrium distribution function at small
deviation from thermal equilibrium.

Understanding the relaxation properties of the one-dimensional
Fermi gas is required for the evaluation of its transport
coefficients, such as thermal conductivity.  At $T\ll\mu$ one can
identify two kinds of thermal conductivity
\cite{matveev_propagation_2018, matveev_thermal_2019,
  samanta_thermal_2019}.  The ordinary thermal conductivity $\kappa$
is controlled by the exponentially rare processes involving
backscattering of particles near the bottom of the band.  It describes
thermal transport at exponentially small frequencies.  At higher
frequencies the thermal transport is described by a different
transport coefficient $\kappa_{\rm ex}$, which is essentially the
thermal conductivity of the gas of elementary excitations of the
system \cite{matveev_propagation_2018}.  Our treatment of the
relaxation of the one-dimensional Fermi gas will enable us to express
$\kappa_{\rm ex}$ in terms of temperature, chemical potential, and
interaction strength.


The paper is organized as follows.  In Sec.~\ref{sec:scattering_rate}
we evaluate the three-particle scattering rates associated with the
two types of processes illustrated in Fig.~\ref{fig:processes}.  In
Sec.~\ref{sec:quasiparticle_decay} we estimate the decay rates of
quasiparticle states with energies of order $T$ and discuss how these
rates scale with the temperature for weak interaction potentials
decaying with the distance as $1/|x|^\gamma$.  In
Sec.~\ref{sec:relaxation} we solve the relaxation problem in the
regime of short-range interactions ($\gamma>5/2$).  To leading order
in small temperature and weak interaction, the corresponding
linearized collision integral is diagonalized exactly in
Appendix~\ref{sec:integral-eq}.  The
spectrum of the relaxation rates is qualitatively different from that
in the spin-polarized system, which is briefly discussed in
Appendix~\ref{sec:spinless}.  We discuss our results and their
implications for the transport coefficients of the one-dimensional
Fermi gas in Sec.~\ref{sec:discussion}.  

\section{Three particle scattering rate}
\label{sec:scattering_rate}

We consider a system of one-dimensional spin-$\frac12$ fermions with
quadratic dispersion $\varepsilon_p=p^2/2m$ and weak two-particle
interaction, which we describe by the Hamiltonian
\begin{equation}
  \label{eq:interaction_hamiltonian}
  \hat V = \frac{1}{2L}\sum_{\substack{p_1,p_2,q\\ \sigma_1,\sigma_2}}
  V(q)c_{p_1+q,\sigma_1}^\dagger c_{p_2-q,\sigma_2}^\dagger
      c_{p_2,\sigma_2}^{}c_{p_1,\sigma_1}^{}.
\end{equation}
Here $L$ is the system size, $V(q)$ is the Fourier transform of the
interaction potential, and $c_{p,\sigma}^{}$ is the operator
annihilating a fermion with momentum $p$ and $z$-component of spin
$\sigma$.

In one dimension the restrictions imposed by conservation of momentum
and energy preclude relaxation by two-particle scattering processes.
Thus the dominant scattering processes involve three particles.
Because the interaction (\ref{eq:interaction_hamiltonian}) couples
only two fermions, the three-particle scattering amplitude must be
obtained in the second order of the perturbation theory in $V(q)$.
Such a calculation was performed in
Ref.~\cite{lunde_three-particle_2007}.  The rate of scattering of
three fermions with momenta $p_1$, $p_2$, $p_3$ and spins $\sigma_1$,
$\sigma_2$, $\sigma_3$ to new states with momenta $p_1'$, $p_2'$,
$p_3'$ and spins $\sigma_1'$, $\sigma_2'$, $\sigma_3'$, respectively,
has the form
\begin{equation}
  \label{eq:rateW}
  \mathcal W_{123}^{1'2'3'}=\frac{2\pi}{\hbar}
  \big|\mathcal{A}_{123}^{1'2'3'}\big|^2\delta(E-E'),
\end{equation}
where $E=\varepsilon_{p_1}+\varepsilon_{p_2}+\varepsilon_{p_3}$ and
$E'=\varepsilon_{p_1}'+\varepsilon_{p_2}'+\varepsilon_{p_3}'$ are the
energies of the three particles before and after the scattering event
and $\mathcal{A}_{123}^{1'2'3'}$ is the scattering matrix element.
The latter can be presented in the form
\begin{eqnarray}
  \label{eq:amplitude}
  \mathcal{A}_{123}^{1'2'3'}&=&\sum_{\pi(1'2'3')}\mathrm{sign}(1'2'3')
  \delta_{\sigma_1,\sigma_1'} \delta_{\sigma_2,\sigma_2'} \delta_{\sigma_3,\sigma_3'}
\nonumber\\
   &&\times
      \big(a_{p_1,p_2}^{p_{a},p_{b}}+a_{p_1,p_3}^{p_{a},p_{c}}+a_{p_2,p_3}^{p_{b},p_{c}}\big)
      \delta_{P,P'}.
\end{eqnarray}
Here the summation is performed over all the permutations of the final
states of the three particles, $P=p_1+p_2+p_3$ and $P'=p_1'+p_2'+p_3'$
are total momenta before and after the scattering event, and
\begin{eqnarray}
  \label{eq:partial_amplitude}
  a_{p_1,p_2}^{p_{a},p_{b}}&=&\frac{1}{L^2}V({p_a-p_1})V({p_b-p_2})
\nonumber\\
   &&\times
      \bigg(\frac{1}{E-\varepsilon_{p_1}-\varepsilon_{p_b}-\varepsilon_{P-p_1-p_b}}
\nonumber\\
   &&\hspace{1em}
   +\frac{1}{E-\varepsilon_{p_2}-\varepsilon_{p_a}-\varepsilon_{P-p_2-p_a}} \bigg).
\end{eqnarray}

In the absence of magnetic field the occupation numbers of all the
states do not depend on the spin.  Thus it will be convenient to sum
the scattering rate (\ref{eq:rateW}) over spin indices and introduce
\begin{equation}
  \label{eq:W_definition}
 W_{p_1,p_2,p_3}^{p_1',p_2',p_3'}=\sum_{\sigma_1,\sigma_2,\sigma_3\atop
   \sigma_1',\sigma_2',\sigma_3'} \mathcal W_{123}^{1'2'3'}.
\end{equation}

Our goal is to evaluate the scattering rate (\ref{eq:W_definition})
assuming that all three fermions are near the Fermi points, see
Fig.~\ref{fig:processes}.  We start with the state described by the
momenta of the three particles and notice that collisions conserve the
total momentum $P=p_1+p_2+p_3$ and energy $E=(p_1^2+p_2^2+p_3^2)/2m$.
Aside from $P$ and $E$, a full description of a state of three
particles requires one additional parameter.  We will denote this
parameter $\alpha$ and introduce it via
\begin{subequations}
\label{eq:Jacobi}
\begin{equation}
\label{eq:Jacobi-initial}
  p_j=\frac{1}{3}P-2\sqrt{\frac{m\mathcal E}{3}}\,
\cos\left(\alpha-\frac{2\pi j}{3}\right),
\quad
j=1,2,3,
\end{equation}
where $\mathcal E=E-P^2/6m$ is the total energy of the three fermions
in the center-of-mass frame.  Thus the state of three particles will
be described by $P$, $\mathcal E$, and $\alpha$.  The momenta of the
three particles after the collision will be similarly parametrized by
$P'$, $\mathcal E'$, and $\alpha'$ according to
\begin{equation}
\label{eq:Jacobi-final}
p_j'=\frac{1}{3}P'-2\sqrt{\frac{m\mathcal E'}{3}}\,
\cos\left(\alpha'-\frac{2\pi j}{3}\right),
\quad
j=1,2,3.
\end{equation}
\end{subequations}
Conservation of momentum and energy implies that the scattering rate
(\ref{eq:W_definition}) has the form
\begin{equation}
  \label{eq:Theta}
  W_{p_1,p_2,p_3}^{p_1',p_2',p_3'}=\Theta\,
  \delta(\mathcal E-\mathcal E')\delta_{P,P'},
\end{equation}
where $\Theta$ is in general a function of $\mathcal E$, $\alpha$, and
$\alpha'$.  (The dependence of $\Theta$ on the total momentum $P$ is
precluded by Galilean invariance.)

The scattering process shown in Fig.~\ref{fig:processes}(a) involves
two fermions near the right Fermi point $p=p_F$ and the third one
near $-p_F$.  In terms of our variables $P$, $\mathcal E$ and
$\alpha$, these conditions translate to
\begin{equation}
  \label{eq:2-1-Jacobi}
  |P-p_F|\lesssim \frac{T}{\mu}\,p_F,
  \quad
  \left|\mathcal E-\frac{8}{3}\,\mu\right|\lesssim T,
  \quad
  |\alpha|\lesssim \frac{T}{\mu},
\end{equation}
where the chemical potential $\mu$ is given by the Fermi energy
$p_F^2/2m$ in the low-temperature limit.  For the processes of
Fig.~\ref{fig:processes}(b), in which all three fermions are near the
right Fermi point, we have
\begin{equation}
  \label{eq:3-0-Jacobi}
  |P-3p_F|\lesssim \frac{T}{\mu}\,p_F,
  \quad
  \mathcal E\lesssim \frac{T^2}{\mu},
  \quad
  -\frac{\pi}{3}<\alpha < \frac{\pi}{3}.
\end{equation}
The estimate of $\mathcal E$ at low temperatures is obtained by
noticing that the typical difference of momenta of the fermions
$|p_1-p_2|\sim\sqrt{m\mathcal E}$ is of the order of $T/v_F$, where
$v_F=p_F/m$ is the Fermi velocity.

If the interaction potential falls off with the distance sufficiently
slowly, the Fourier transform $V(q)$ is not analytic at $q=0$.  For
example, for the Coulomb interaction, $V(q)\propto\ln (1/|q|)$.
Relaxation of the Fermi gas with such long-range interactions has a
number special features, which we leave for future study.  In the
following we assume that the range of interactions between fermions is
short.  The exact criterion for a potential to be considered
short-range depends on the particular result and will be discussed
below.  An interaction potential that decays exponentially at large
distances corresponds to $V(q)$ that is analytic at $q=0$.  This will
be sufficient to classify such potentials as short-range, but in
practice exponential decay will not be required.

In the case of a short-range potential, the general result for the
three-particle scattering rate $W_{p_1,p_2,p_3}^{p_1',p_2',p_3'}$ given by
Eqs.~(\ref{eq:rateW})--(\ref{eq:W_definition}) can be simplified for
the two types of processes that dominate relaxation at low
temperatures, see Fig.~\ref{fig:processes}.  For the process of
Fig.~\ref{fig:processes}(a) we use the simplification
(\ref{eq:2-1-Jacobi}) and find the scattering rate in the form
(\ref{eq:Theta}) with
\begin{equation}
  \Theta=\frac{9\pi}{L^4}
  \frac{\hbar^3\Lambda^2 }{m^2}
  \frac{\alpha^2+\alpha'^2}{(\alpha^2-\alpha'^2)^2},
\label{eq:W21}
\end{equation}
where $\Lambda$ is a dimensionless parameter defined as
\begin{equation}
  \label{eq:Lambda}
  \Lambda=\frac{V(0)V(2p_F)-V(2p_F)^2-2p_F V(0)V'(2p_F)}{(\hbar v_F)^2}. 
\end{equation}
The result (\ref{eq:W21}) is applicable as long as $V(0)$ is well
defined.  For interactions that fall off with the distance as
$1/|x|^\gamma$ this requires $\gamma>1$.

For the scattering processes of Fig.~\ref{fig:processes}(b) we have
$\mathcal E\ll\mu$, see Eq.~(\ref{eq:3-0-Jacobi}), which enables one
to simplify the general expression for the scattering rate
$W_{p_1,p_2,p_3}^{p_1',p_2',p_3'}$ given by
Eqs.~(\ref{eq:rateW})--(\ref{eq:W_definition}) to the form
(\ref{eq:Theta}) with
\begin{equation}
  \Theta=\frac{1458\pi m^2}{\hbar L^4}
  [V(0)V''(0)]^2
  \frac{1-\cos(3\alpha)\cos(3\alpha')}
  {\left[\cos(3\alpha)-\cos(3\alpha')\right]^2}.
\label{eq:W30}
\end{equation}
The applicability of this expression is limited to interaction
potentials for which the second derivative of the Fourier transform
$V''(q)$ is well defined at $q=0$.  For interactions that fall off as
$1/|x|^\gamma$ this requires $\gamma>3$.

It is instructive to consider a special case of $V(q)=\rm const$,
which corresponds to the interaction of fermions in the form
$U(x)\propto\delta(x)$.  The model of spin-$\frac12$ fermions with
interaction of this type was studied by Gaudin and Yang and shown to
be integrable \cite{gaudin_systeme_1967, yang_exact_1967}.  This
property implies that no scattering of elementary excitations is
allowed \cite{sutherland_beautiful_2004}.  Substitution of
$V(q)=\rm const$ into Eqs.~(\ref{eq:W21}) and (\ref{eq:W30}) indeed
yields $\Theta=0$.  More generally, integrability should result in a
vanishing scattering amplitude (\ref{eq:amplitude}) for
$V(q)=\rm const$. This was verified in
Ref.~\cite{lunde_three-particle_2007}.

\section{Decay of quasiparticle states}
\label{sec:quasiparticle_decay}

As a first step toward understanding relaxation of the one-dimensional
Fermi gas we estimate the decay rates of quasiparticles due to the
three-particle scattering processes.  For a quasiparticle of momentum
$p$ the decay rate is given by
\begin{eqnarray}
  \label{eq:qp-decay-1}
  \frac{1}{\tau}&=&\frac12\!
  \sum_{p_1,p_2,p_3\atop{p_1',p_2',p_3'}}\hspace{-1em}
  \, W_{p_1,p_2,p_3}^{p_1',p_2',p_3'}
  \theta(p_2-p_3)\theta(p_1'-p_2')\theta(p_2'-p_3')\delta_{p,p_1}
  \nonumber\\
  &&\times
  n_{p_2} n_{p_3}(1-n_{p_1'}) (1-n_{p_2'}) (1-n_{p_3'}).
\end{eqnarray}
Here the unit step function $\theta(x)$ is used to limit the
summations to distinct sets of momenta before and after scattering,
and 1/2 compensates for the summation over the spin of the initial
particle included in Eq.~(\ref{eq:W_definition}).  To estimate the
rate, we convert the sum to an integral and substitute the general
form (\ref{eq:Theta}) of the scattering rate.  This yields
\begin{eqnarray}
  \label{eq:qp-decay-2}
  \frac{1}{\tau}&=&
  \frac{L^4m^2}{384\pi^4\hbar^4}\!
  \iiint \!d\mathcal E\,d\alpha\,d\alpha'\,
  \Theta\,
\nonumber\\
  &&\times
  n_{p_2} n_{p_3}(1-n_{p_1'}) (1-n_{p_2'}) 
  (1-n_{p_3'}).
\end{eqnarray}
Here we transformed the integral to the  variables (\ref{eq:Jacobi}) using
\begin{equation}
  \label{eq:Jacobian}
  dp_1dp_2dp_3=\frac{m}{\sqrt3}\,dP\,d\mathcal E\, d\alpha.
\end{equation}
The Fermi occupation numbers, to which one should substitute the
expressions for momenta using Eq.~(\ref{eq:Jacobi}), effectively limit
the range of integration in Eq.~(\ref{eq:qp-decay-2}).

Assuming that the quasiparticle of interest has the energy within $T$
from the Fermi level, its decay is controlled by the two processes
shown in Fig.~\ref{fig:processes}.  We start with the process shown in
Fig.~\ref{fig:processes}(a) and substitute into
Eq.~(\ref{eq:qp-decay-2}) the expression (\ref{eq:W21}) for $\Theta$.
This yields
\begin{equation}
  \label{eq:decay-21-general}
  \frac{1}{\tau}\sim
  \frac{\Lambda^2}{\hbar}
  \int\limits_{\left|\mathcal E -\frac{8\mu}{3}\right|\lesssim T}
  \hspace{-1.5em}d\mathcal E
  \hspace{1em}
  \iint\limits_{|\alpha|,|\alpha'|\lesssim\frac{T}{\mu}}
  \hspace{-1em}
  d\alpha\,d\alpha'
  \frac{\alpha^2+\alpha'^2}{(\alpha^2-\alpha'^2)^2},
\end{equation}
with the ranges of integrations controlled by the omitted Fermi
occupation numbers, see Eq.~(\ref{eq:2-1-Jacobi}).  Ignoring for the
moment the singularity at $\alpha=\pm\alpha'$, we find that the
integral over $\alpha$ and $\alpha'$ is of order unity, while the
integral over $\mathcal E$ is of the order of $T$.  We therefore
conclude that the processes of Fig.~\ref{fig:processes}(a) result in
the relaxation rate of the order of
\begin{equation}
  \label{eq:decay-21-result}
  \frac{1}{\tau_a}=
  \frac{\Lambda^2}{\hbar}\,T.
\end{equation}
Similarly, for the processes of Fig.~\ref{fig:processes}(b)
substitution of Eq.~(\ref{eq:W30}) into Eq.~(\ref{eq:qp-decay-2})
yields
\begin{eqnarray}
  \label{eq:decay-30-short-general}
  \frac{1}{\tau}&\sim&
  \frac{m^4}{\hbar^5}
  [V(0)V''(0)]^2
  \int\limits_{0<\mathcal E \lesssim \frac{T^2}{\mu}}
  \hspace{-1em}
  d\mathcal E
                       \nonumber\\
  &&\times
  \iint\limits_{|\alpha|,|\alpha'|<\frac{\pi}{3}}
  d\alpha\,d\alpha'
  \frac{1-\cos(3\alpha)\cos(3\alpha')}
  {\left[\cos(3\alpha)-\cos(3\alpha')\right]^2}.
\end{eqnarray}
The corresponding relaxation rate is
\begin{equation}
  \label{eq:decay-30-short-result}
  \frac{1}{\tau_b}=
  \frac{m^4}{\hbar^5\mu}
  [V(0)V''(0)]^2
  T^2.
\end{equation}

Our estimates (\ref{eq:decay-21-result}) and
(\ref{eq:decay-30-short-result}) should be understood as follows.  The
quasiparticle decay rates (\ref{eq:decay-21-general}) and
(\ref{eq:decay-30-short-general}) diverge due to the singularities at
$\alpha=\pm\alpha'$.  One can see from Eq.~(\ref{eq:Jacobi}) that
these divergences emerge as a result of scattering processes for which
the fermion with momentum $p$ scatters to a state with momentum $p'$
approaching $p$.  Within our perturbative treatment, in the lowest
order in interaction strength, the decay rate is infinite.  On the
other hand, an infinitesimal change of momentum of the fermion from
$p$ to $p'$ has little effect on the observable quantities.  In the
next section we will see that the evolution of the fermion
distribution function is not affected by these singularities.  Thus
the expressions (\ref{eq:decay-21-result}) and
(\ref{eq:decay-30-short-result}) give the order of magnitude estimates
of the relaxation rates associated with the processes shown in
Fig.~\ref{fig:processes}.

In the above calculation we assumed that the fermion in the state with
momentum $p$ had the energy $\varepsilon_p$ near the Fermi energy,
$|\varepsilon_p-\mu|\sim T$.  Decay of quasiparticles with energies
larger than temperature in an electron gas with Coulomb interactions
was studied in Ref.~\cite{karzig_energy_2010}.  Only the processes of
the type shown in Fig.~\ref{fig:processes}(a) were considered.  At
energies of order $T$ the corresponding results of
Ref.~\cite{karzig_energy_2010} are consistent with our estimate
(\ref{eq:decay-21-result}) provided the logarithmic singularity of
$V(q)\propto\ln (1/|q|)$ is properly cut off.

Comparison of the expressions (\ref{eq:decay-21-result}) and
(\ref{eq:decay-30-short-result}) shows that at low temperature
relaxation is dominated by the processes of
Fig.~\ref{fig:processes}(a).  This conclusion holds for sufficiently
short-range interactions, such that $V''(0)$ is well defined.  In the
case of a potential that falls off as a power-law $|x|^{-\gamma}$ at
large distances, this condition requires $\gamma>3$.  For $1<\gamma<3$
the temperature dependence of the rate $\tau_b^{-1}$ can be obtained
as follows.  The Fourier transform of the interaction potential $V(q)$
is well defined at $q=0$ for $\gamma>1$.  However its second
derivative diverges at $q\to0$ as $V''(q)\propto |q|^{\gamma-3}$.  For
the process shown in Fig.~\ref{fig:processes}(b) the typical
difference of momenta in the argument of $V$ in
Eq.~\eqref{eq:partial_amplitude} is of the order of $T/v_F$.  Thus one
can obtain the temperature dependence of $\tau_b^{-1}$ by substituting
$V''(T/v_F)\propto T^{\gamma-3}$ for $V''(0)$ in
Eq.~(\ref{eq:decay-30-short-result}).  This yields
\begin{equation}
  \label{eq:tau_b_estimate}
  \frac{1}{\tau_b}\propto T^{2\gamma-4}.
\end{equation}
At $T\to0$ the above rate is negligible compared with
$1/\tau_a\propto T$ if $\gamma>5/2$.  Conversely, for $1<\gamma<5/2$
we expect relaxation to be dominated by the processes of
Fig.~\ref{fig:processes}(b).

\section{Relaxation of the distribution function}
\label{sec:relaxation}

We now consider how the one-dimensional Fermi gas relaxes to its
equilibrium state.  The latter is described by the occupation numbers
of the different momentum states in the Fermi-Dirac form
\begin{equation}
  \label{eq:Fermi}
  n_p^{(0)}=\frac{1}{e^{(\varepsilon_p-\mu)/T}+1}.
\end{equation}
The evolution of the occupation numbers $n_p$ toward the equilibrium
values (\ref{eq:Fermi}) due to the three-particle collisions is
described by the following collision integral
\begin{eqnarray}
  \label{eq:dotn_p}
  \hspace{-2em}
  \dot n_{p}&=&-\frac12\!
  \sum_{p_1,p_2,p_3\atop{p_1',p_2',p_3'}}\hspace{-1em}
  \, W_{p_1,p_2,p_3}^{p_1',p_2',p_3'}
  \theta(p_2-p_3)\theta(p_1'-p_2')\theta(p_2'-p_3')
\nonumber\\
              &&\ \times \delta_{p,p_1}
  \left[n_{p_1}n_{p_2}n_{p_3}(1-n_{p_1'})(1-n_{p_2'})(1-n_{p_3'})\right.
\nonumber\\
              &&\ \left.
  -(1-n_{p_1})(1-n_{p_2})(1-n_{p_3})n_{p_1'}n_{p_2'}n_{p_3'}\right].
\end{eqnarray}
Here we again limit the summation to non-equivalent sets of initial as
well as final momenta.

For a system near thermal equilibrium it is convenient to present
occupation numbers in the form
\begin{equation}
  \label{eq:phi_definition}
  n_p=n_p^{(0)}+g_p\phi_p,
\end{equation}
where
\begin{equation}
  \label{eq:g_p}
  g_p=\sqrt{n_p^{(0)}\Big(1-n_p^{(0)}\Big)}
     =\frac{1}{2\cosh\frac{\varepsilon_p-\mu}{2T}}.
\end{equation}
We then substitute Eq.~(\ref{eq:phi_definition}) into
Eq.~(\ref{eq:dotn_p}), linearize in small $\phi_p$ and obtain
\begin{equation}
  \label{eq:dotphi_p}
  \dot \phi_{p}=-\widehat W\phi_p,
\end{equation}
where the linearized collision integral $\widehat W$ is defined by
\begin{eqnarray}
  \label{eq:W_operator}
   \hspace{-.7em}
  \widehat W \phi_{p}&=&\frac12\!
  \sum_{p_1,p_2,p_3\atop{p_1',p_2',p_3'}}\hspace{-1em}
  \, W_{p_1,p_2,p_3}^{p_1',p_2',p_3'}
  \theta(p_2-p_3)\theta(p_1'-p_2')\theta(p_2'-p_3')
\nonumber\\
              &&\times \delta_{p,p_1}
  g_{p_2}g_{p_3}g_{p_1'}g_{p_2'}g_{p_3'}
\nonumber\\
              &&\times
                 \left(\frac{\phi_{p_1}}{g_{p_1}}+\frac{\phi_{p_2}}{g_{p_2}}
                 +\frac{\phi_{p_3}}{g_{p_3}}
                 -\frac{\phi_{p_1'}}{g_{p_1'}}-\frac{\phi_{p_2'}}{g_{p_2'}}
                 -\frac{\phi_{p_3'}}{g_{p_3'}} \right).
\end{eqnarray}
The problem of the relaxation of the system to thermodynamic
equilibrium has now been reduced to solving Eq.~(\ref{eq:dotphi_p}).
Since $\widehat W$ is a real symmetric linear integral operator, one
can, in principle, solve the eigenvalue problem
\begin{equation}
  \label{eq:eigenvalue_problem}
  \widehat W \phi_p^{(l)}=\frac{1}{\tau_l^{}} \phi_p^{(l)}
\end{equation}
and obtain real eigenvalues $\tau_l^{-1}$.  A general solution of
Eq.~(\ref{eq:dotphi_p}) is then obtained as a linear combination
\begin{equation}
  \label{eq:relaxation_solutions}
  \phi_p(t)=\sum_lC_l\,e^{-t/\tau_l}\phi_p^{(l)}.
\end{equation}
Thus the eigenvalues defined by Eq.~(\ref{eq:eigenvalue_problem}) are
the relaxation rates associated with modes $\phi_p^{(l)}$.

Our goal is to study relaxation of the one-dimensional Fermi gas at
low temperatures $T\ll\mu$.  As discussed above, the relaxation is
dominated by the three-particle processes shown in
Fig.~\ref{fig:processes}.  We limit ourselves to the relatively
short-range interactions that fall off faster than $1/|x|^{5/2}$.  As
we discussed in Sec.~\ref{sec:quasiparticle_decay}, for such
interactions relaxation is dominated by the processes shown in
Fig.~\ref{fig:processes}(a).  Thus from now on the processes of
Fig.~\ref{fig:processes}(b) will be neglected.

An important feature of the process shown in
Fig.~\ref{fig:processes}(a) is that while all the momenta of the
initial and final states of the fermions measured from the nearest
Fermi point are of the order or $T/v_F$, the difference of momenta
$|p_3-p_3'|$ is much smaller than $T/v_F$.  Indeed, using
Eqs.~(\ref{eq:Jacobi}) and (\ref{eq:2-1-Jacobi}), we find
\begin{equation}
  \label{eq:delta_p3_estimate}
  |p_3-p_3'|\simeq \frac{2p_F}{3}\big|\alpha^2-\alpha'^2\big|
  \lesssim\frac{T^2}{v_F\mu}\ll \frac{T}{v_F}.
\end{equation}
This feature can be understood as follows.  The possible values of
momenta of the three particles before and after collision are
restricted by the momentum and energy conservation laws.  At low
temperature the energy spectrum of the particles near the Fermi points
is approximately linear,
\begin{equation}
  \label{eq:linearization}
  \varepsilon_p=\frac{p^2}{2m}\simeq \mu+v_F(|p|-p_F).
\end{equation}
In this approximation, any choice of momenta   $p_1$, $p_2$, $p_1'$,
and $p_2'$ such that $p_1+p_2=p_1'+p_2'$ guarantees that
$\varepsilon_{p_1}+\varepsilon_{p_2}
=\varepsilon_{p_1'}+\varepsilon_{p_2'}$.  Thus both momentum and
energy are conserved if $p_3=p_3'$.  A small quadratic correction to
the energy in Eq.~(\ref{eq:linearization}) results in small
$|p_3-p_3'|$, see Eq.~(\ref{eq:delta_p3_estimate}).

Nonlinearity of the energy spectrum must be taken into account when
solving the quantum-mechanical problem of evaluation of the
three-particle scattering rate, see Sec.~\ref{sec:scattering_rate}.
Linearization of the spectrum at that stage would lead to singular
scattering rates.  On the other hand, the collision integral in both
its original and linearized forms (\ref{eq:dotn_p}) and
(\ref{eq:W_operator}) takes finite values when the spectrum approaches
linear form (\ref{eq:linearization}).  This procedure is appropriate
only for studying the relaxation of the system in the leading order at
low temperature \cite{onefootnote}.
Because in this approximation $p_3=p_3'$, the distribution function of
the particles near the left Fermi point in Fig.~\ref{fig:processes}(a)
remains unchanged.  Thus, to leading order in $T\ll\mu$ the subsystems
of right- and left-moving particles relax independently of each other.

We now substitute Eqs.~(\ref{eq:Theta}) and (\ref{eq:W21}) into the
definition (\ref{eq:W_operator}) of the operator $\widehat W$ and use
Eqs.~(\ref{eq:Jacobi}) and (\ref{eq:Jacobian}) to convert the sum into
an integral over $P$, $P'$, $\mathcal E$, $\mathcal E'$, $\alpha$, and
$\alpha'$.  Assuming $p_3=p_3'$, the integral over the first four of
these variables is straightforward and yields
\begin{eqnarray}
  \label{eq:W_operator_linearization}
  \widehat W \phi_{p}&=&\frac{3}{16\pi^3\tau_a}
  \iint d\alpha\, d\alpha'
  \frac{\alpha^2+\alpha'^2}{(\alpha^2-\alpha'^2)^2}\,
  g_{p_2}g_{p_1'}g_{p_2'}
\nonumber\\
              &&\times
                 \left(\frac{\phi_{p}}{g_{p}}+\frac{\phi_{p_2}}{g_{p_2}}
                 -\frac{\phi_{p_1'}}{g_{p_1'}}-\frac{\phi_{p_2'}}{g_{p_2'}}
                 \right).
\end{eqnarray}
Here $\tau_a$ is defined by Eq.~(\ref{eq:decay-21-result}), the
spectrum $\varepsilon_p$ in the definition (\ref{eq:g_p}) of $g_p$ is
linearized according to Eq.~(\ref{eq:linearization}),
\begin{equation}
  \label{eq:g_p_linearized}
  g_p=\frac{1}{2\cosh \frac{v_F(|p|-p_F)}{2T}},
\end{equation}
and the momenta
\begin{equation}
  \label{eq:momenta_linearized}
  p_2=p+\frac{4p_F}{\sqrt3}\,\alpha,
  \quad
  p_{1,2}'=p+\frac{2p_F}{\sqrt3}\,\big(\alpha\mp\alpha'\big).
\end{equation}
are evaluated to linear order in $T$ using Eqs.~(\ref{eq:Jacobi}) and
(\ref{eq:2-1-Jacobi}).  Given that $g_p$ falls off exponentially away
from $p=p_F$, the integrals over $\alpha$ and $\alpha'$ in
Eq.~(\ref{eq:W_operator_linearization}) should be taken from $-\infty$
to $+\infty$.

Similarly to the expression (\ref{eq:decay-21-general}) for the
quasiparticle decay rate, the integrand of
Eq.~(\ref{eq:W_operator_linearization}) contains a factor
$1/(\alpha^2-\alpha'^2)^2$, which diverges at $\alpha=\pm\alpha'$.
However, one can easily see from Eq.~(\ref{eq:momenta_linearized})
that the expression in the second line of
Eq.~(\ref{eq:W_operator_linearization}) vanishes at
$\alpha=\pm\alpha'$.  Thus the integrand is only singular as
$1/(\alpha^2-\alpha'^2)$, resulting in a finite integral that should
be treated as a principal value.

The eigenvalue problem (\ref{eq:eigenvalue_problem}) with $\widehat W$
defined by
Eqs.~(\ref{eq:W_operator_linearization})--(\ref{eq:momenta_linearized})
can be solved exactly, see Appendix~\ref{sec:integral-eq}.  The
eigenvalues and eigenfunctions are
\begin{eqnarray}
  \label{eq:relaxation_rates}
  \frac{1}{\tau_l}&=&\frac{3}{32\pi^3\tau_a}
  \times
  \left\{
  \begin{array}[c]{ll}
    \displaystyle
    \sum_{j=1}^{l}\frac{1}{j}, &\mbox{for even $l$,}
    \\[4ex]
    \displaystyle
    \sum_{j=1}^{l}\frac{1}{j}-\frac{2}{l(l+1)}, &\mbox{for odd $l$,}
  \end{array}
    \right.
\\[2ex]
  \label{eq:eigenfunctions-right}
  \phi_p^{(l)}&=&\theta(p) B_l\left(\frac{v_F(p-p_F)}{\pi T}\right)g_p.
\end{eqnarray}
Here $l=0,1,2,\ldots$, the rate $\tau_a^{-1}$ is defined by
Eq.~(\ref{eq:decay-21-result}), and $B_l(u)$ are modified Bateman
polynomials \cite{bateman_properties_1933, bateman_polynomial_1934,
  threefootnote} defined by
\begin{equation}
  \label{eq:Bateman_modified}
  B_l(u)
  =\frac{i^l}{\pi}\cosh\frac{\pi u}{2}
  \int_{-\infty}^{+\infty} dx\,e^{-iux}\frac{P_l^{}(\tanh x)}{\cosh x},
\end{equation}
where $P_l^{}(y)$ are the Legendre polynomials.  In particular,
\begin{equation}
  \label{eq:Bateman_low_l}
  B_0(u)=1,
  \quad
  B_1(u)=u,
  \quad
  B_2(u)=\frac{3u^2-1}{4}.
\end{equation}
The step function $\theta(p)$ in Eq.~(\ref{eq:eigenfunctions-right})
accounts for the fact that in the linearized spectrum approximation
only the right-moving particles are scattered in
Fig.~\ref{fig:processes}(a).  In Eq.~(\ref{eq:eigenfunctions-right})
we omitted the normalization factor, which can be restored with the
help of Eq.~(\ref{eq:Phi_l}).

In addition to the processes illustrated in
Fig.~\ref{fig:processes}(a) there are similar ones that involve two
particles near the left Fermi point and one particle near the right
one.  These processes equilibrate the left-moving particles.
Inversion symmetry dictates that the relaxation rates are again given
by Eq.~(\ref{eq:relaxation_rates}) with the relaxation modes
\begin{equation}
  \label{eq:eigenfunctions-left}
  \phi_p^{(l)}=\theta(-p) B_l\left(-\frac{v_F(p+p_F)}{\pi T}\right)g_p.
\end{equation}

For $l=0$ and 1 the relaxation rates (\ref{eq:relaxation_rates})
vanish.  The corresponding eigenfunctions
(\ref{eq:eigenfunctions-right}) are
\begin{equation}
  \label{eq:zero_modes}
  \phi_p^{(0)}=\theta(p) g_p,
  \qquad
  \phi_p^{(1)}=\theta(p) \frac{v_F(p-p_F)}{\pi T}\,g_p.
\end{equation}
Indeed, from Eq.~(\ref{eq:W_operator_linearization}) one immediately
obtains $\widehat W \phi_p^{(0)}=0$, as the expression in parentheses
vanishes.  The deviation of the distribution function from the
equilibrium form (\ref{eq:Fermi}) described by $\phi_p^{(0)}$
corresponds to a small change of the chemical potential $\mu$.  Thus
this zero mode reflects the conservation of the number of particles
near the right Fermi point.  To verify that
$\widehat W \phi_p^{(1)}=0$, one should keep in mind that
$p+p_2=p_1'+p_2'$, which follows immediately from
Eq.~(\ref{eq:momenta_linearized}).  The latter condition is satisfied
automatically for the linearized spectrum (\ref{eq:linearization})
because in this case $p_3=p_3'$.  Alternatively, the same condition
can be interpreted as conservation of energy of the two particles near
the right Fermi point.  Correspondingly, the deviation of the
distribution function from the equilibrium form (\ref{eq:Fermi})
described by $\phi_p^{(1)}$ can be interpreted as a result of a small
change of temperature.

\section{Discussion of the results}
\label{sec:discussion}

In this paper we have studied the relaxation of a gas of
one-dimensional spin-$\frac12$ fermions at low temperatures.  We
focused on the case of small deviations of the distribution function
from the equilibrium Fermi-Dirac form (\ref{eq:Fermi}).  This enabled
us to linearize the collision integral and obtain the spectrum of
relaxation rates (\ref{eq:relaxation_rates}) in terms of the
interaction potential and temperature.  To leading orders in small
temperature and weak interactions the result
(\ref{eq:relaxation_rates}) is exact.

The relaxation rates (\ref{eq:relaxation_rates}) scale linearly with
the temperature, see Eq.~(\ref{eq:decay-21-result}).  This conclusion
is consistent with the expectation based on the earlier results for
the quasiparticle energy relaxation rate in a one-dimensional electron
gas with Coulomb interactions \cite{karzig_energy_2010}.  Unlike the
authors of Ref.~\cite{karzig_energy_2010}, we considered both
three-particle processes shown in Fig.~\ref{fig:processes}.  We showed
that the scattering processes of Fig.~\ref{fig:processes}(b), which
were neglected in Ref.~\cite{karzig_energy_2010}, give subleading
contribution to the relaxation rate provided that the interaction
between the fermions falls off with the distance faster than
$1/|x|^{5/2}$.  An important example of such a one-dimensional Fermi
system is the electron gas in a quantum wire with a metal gate
parallel to it, in which case interactions fall off as $1/|x|^3$.  The
expected temperature dependence for more slowly decaying potentials is
given by Eq.~(\ref{eq:tau_b_estimate}).

Our result (\ref{eq:relaxation_rates}) predicts a discrete spectrum of
the relaxation rates.  It is instructive to compare this behavior with
the case of spin-polarized one-dimensional Fermi gas.  The linearized
collision integral analogous to
Eq.~(\ref{eq:W_operator_linearization}) was obtained in
Ref.~\cite{matveev_thermal_2019}.  It can be diagonalized numerically,
see Appendix~\ref{sec:spinless}.  Importantly, the spectrum of
relaxation rates is continuous.  The relaxation modes are
qualitatively different as well.  Specifically, each mode of the
continuous spectrum has a singularity at a certain value of momentum
and can be associated with decay of a particular quasiparticle state.
Continuous spectrum and singularities in relaxation modes were also
obtained in other systems of spin-polarized fermions
\cite{degottardi_equilibration_2019, lin_thermalization_2013}.  In
contrast, our results (\ref{eq:eigenfunctions-right}) and
(\ref{eq:eigenfunctions-left}) show smooth analytic behavior as a
function of momentum.


Our expressions (\ref{eq:eigenfunctions-right}) and
(\ref{eq:eigenfunctions-left}) for the relaxation modes contain step
functions $\theta(\pm p)$, which limit the ranges of momentum to
either positive or negative values.  This should not be considered to
be a singularity as a function of momentum as our approach is limited
to fermion states near the Fermi points, where the linearization
(\ref{eq:linearization}) of the spectrum is justified.  Given the
inversion symmetry of the problem it is natural to introduce even and
odd modes
\begin{eqnarray}
  \label{eq:eigenfunctions-even}
  \phi_p^{(l,+)}&=& B_l\left(\frac{v_F(|p|-p_F)}{\pi T}\right)g_p,
\\
  \label{eq:eigenfunctions-odd}
  \phi_p^{(l,-)}&=&B_l\left(\frac{v_F(|p|-p_F)}{\pi T}\right)g_p\,{\rm sgn\,}p.
\end{eqnarray}
In the approximation of linearized spectrum the modes $\phi_p^{(l,+)}$
and $\phi_p^{(l,-)}$ have the same relaxation rate
(\ref{eq:relaxation_rates}) for any given $l$.  We expect the main
effect of the spectral curvature to be a small in $T/\mu$ splitting of
the degeneracies of relaxation rates of the even and odd modes.


The relaxation properties of the one-dimensional Fermi gas determine
its transport coefficients, such as the thermal conductivity and
viscosity.  The thermal conductivity of one-dimensional systems of
spinless fermions has been recently studied in
Refs.~\cite{matveev_thermal_2019, samanta_thermal_2019}.  The dc
thermal conductivity $\kappa$ of these systems is controlled by the
processes involving exponentially weak backscattering of particles
near the bottom of the band.  At frequencies above an exponentially
small value $\omega^*\propto \exp(-\mu/T)$, the backscattering
processes are negligible, and the thermal transport is controlled by
the thermal conductivity $\kappa_{\rm ex}$ of the gas of elementary
excitations \cite{matveev_propagation_2018}.  A relation between
$\kappa_{\rm ex}$ and the solutions of the relaxation problem has the
form \cite{matveev_thermal_2019, twofootnote}
\begin{equation}
  \label{eq:kappa_ex_general}
  \kappa_{\rm ex}=\frac{1}{2m^4T^2}\sum_s \tau_s
  \frac{\langle \phi_p^{(s)}|\psi_p \rangle^2}
  {\langle \phi_p^{(s)}|\phi_p^{(s)} \rangle}.
\end{equation}
Here the summation is over all the eigenmodes of the relaxation
problem with nonvanishing rates $\tau_s^{-1}$, the inner product is
defined as
\begin{equation}
  \label{eq:inner_product}
  \langle \alpha_p|\beta_p\rangle=\int_{-\infty}^{+\infty}
  \frac{dp}{2\pi\hbar}\alpha_p\beta_p,
\end{equation}
and $\psi_p$ is given by
\begin{equation}
  \label{eq:nu_p}
  \psi_p=3p_F
  \bigg[
  (|p|-p_F)^2-\frac{\pi^2T^2}{3v_F^2}
  \bigg]g_p\,{\rm sgn\,}p.
\end{equation}
Interestingly, $\psi_p$ coincides with $\phi_p^{(2,-)}$ given by
Eq.~(\ref{eq:eigenfunctions-odd}) up to a momentum-independent factor
$4\pi^2T^2m/v_F$, see Eq.~(\ref{eq:Bateman_low_l}).  Since the
eigenmodes $\phi_p^{(s)}$ are orthogonal to each other, only the term
with $s=\{2,-\}$ in the sum in Eq.~(\ref{eq:kappa_ex_general}) gives a
nonvanishing contribution.  This greatly simplifies the evaluation of
$\kappa_{\rm ex}$, which yields
\begin{equation}
  \label{eq:kappa_ex_result}
  \kappa_{\rm ex}=\frac{2\pi^3}{5}\frac{T^3v_F\tau_2^{}}{\hbar\mu^2}
  =\frac{128 \pi^6}{45}\frac{T^2 v_F}{\Lambda^2\mu^2},
\end{equation}
where we applied Eqs.~(\ref{eq:relaxation_rates}) and
(\ref{eq:decay-21-result}).  The result (\ref{eq:kappa_ex_result})
differs dramatically from $\kappa_{\rm ex}\propto T^{-4}$ in the case
of spinless fermions \cite{matveev_thermal_2019}.  This is due to the
slow relaxation of the spinless system, $\tau^{-1}\propto T^7$
\cite{imambekov_one-dimensional_2012, arzamasovs_kinetics_2014,
  protopopov_relaxation_2014, matveev_thermal_2019}, compared to
$\tau^{-1}\propto T$ for spin-$\frac12$ fermions.  Equation
(\ref{eq:kappa_ex_result}) gives the thermal conductivity of the
one-dimensional Fermi gas in a broad range of frequencies below
$\tau_a^{-1}$ given by Eq.~(\ref{eq:decay-21-result})
\cite{fourfootnote}.

An expression similar to Eq.~(\ref{eq:kappa_ex_general}) can be
obtained for the bulk viscosity $\zeta$ of a one-dimensional spinless
quantum liquid \cite{matveev_viscous_2017}.  Unlike thermal
conductivity, $\zeta$ is controlled by the relaxation modes that are
even with respect to inversion.  In fact, for the bulk viscosity of a
spinless system the analog of $\psi_p$ in
Eq.~(\ref{eq:kappa_ex_general}) is proportional to $\phi_p^{(2,+)}$
defined by Eq.~(\ref{eq:eigenfunctions-even}).  Generalization of the
treatment of bulk viscosity in Ref.~\cite{matveev_viscous_2017} to
systems with spins is not entirely straightforward and will be
discussed elsewhere.

Given the special role that the modes $\phi_p^{(2,+)}$ and
$\phi_p^{(2,-)}$ play in the evaluation of the transport coefficients,
it is worth discussing how the corresponding relaxation rate behaves
for the long range interactions.  As we saw in
Sec.~\ref{sec:quasiparticle_decay}, for interaction potentials that
fall off at $x\to\infty$ as $1/|x|^\gamma$ with $\gamma<5/2$, the
scattering processes of Fig.~\ref{fig:processes}(b) dominate the
relaxation of the Fermi gas.  These processes obey conservation laws
of the number of right-moving particles, their momentum, and energy.
Thus the corresponding collision integral must have three zero modes.
It is easy to see from Eq.~(\ref{eq:W_operator}) that these three
modes are $\phi_p^{(0)}$, $\phi_p^{(1)}$, and $\phi_p^{(2)}$ defined
by Eqs.~(\ref{eq:eigenfunctions-right}) and (\ref{eq:Bateman_low_l}).
Symmetry requires that the modes $\phi_p^{(0)}$, $\phi_p^{(1)}$, and
$\phi_p^{(2)}$ defined by Eq.~(\ref{eq:eigenfunctions-left}) are also
zero modes of the collision integral due to the processes involving
three particles on the same branch.  We therefore conclude that the
even and odd combinations $\phi_p^{(2,+)}$ and $\phi_p^{(2,-)}$ are
not affected by the processes of Fig.~\ref{fig:processes}(b) and
remain eigenfunctions of the linearized collision integral even for
$\gamma<5/2$.  Additionally, our result for $\tau_2^{-1}$ obtained
from Eq.~(\ref{eq:relaxation_rates}) and the result
(\ref{eq:kappa_ex_result}) for $\kappa_{\rm ex}$ remain unchanged for
long-range interactions with $1<\gamma<5/2$.  In the important case of
Coulomb interaction $e^2/|x|$ with a short distance cutoff $w$, which
corresponds to $\gamma=1$, this result is still valid if one
substitutes $\Lambda=(2e^2/\hbar v_F)^2 \ln(p_Fw/\hbar) \ln(\mu/T)$,
cf.\ Ref.~\cite{karzig_energy_2010}.


In this paper the interactions between fermions are treated in the
lowest order of the perturbation theory.  This has enabled us to
ignore the Luttinger liquid effects that develop in interacting
one-dimensional systems at $T\to0$, such as spin-charge separation.
The latter means that instead of quasiparticles and quasiholes with
Fermi statistics the elementary excitations of the system are two
types bosons, in the charge and spin sectors, propagating at different
velocities.  Luttinger liquid effects can be neglected if the
interactions are sufficiently weak compared with the typical energy of
the quasiparticles \cite{karzig_energy_2010}, which in our case is the
temperature.  This results in the condition $p_FV(0)/\hbar\ll T$.

\begin{acknowledgments}

  The authors are grateful to Wade DeGottardi for helpful comments.
  Work at Argonne National Laboratory was supported by the
  U.S. Department of Energy, Office of Science, Basic Energy Sciences,
  Materials Sciences and Engineering Division.  Work at Laboratoire de
  Physique Th\'{e}orique was supported in part by the EUR grant NanoX
  ANR-17-EURE-0009 in the framework of the ``Programme des
  Investissements d’Avenir.''

\end{acknowledgments}

\appendix

\section{Solution of the eigenvalue problem
  (\ref{eq:eigenvalue_problem})}
\label{sec:integral-eq}

Here we solve the eigenvalue problem (\ref{eq:eigenvalue_problem})
with the operator $\widehat W$ defined by
Eqs.~(\ref{eq:W_operator_linearization})--(\ref{eq:momenta_linearized}).
We start by introducing dimensionless variables $u$, $w$, and $w'$ via
\begin{widetext}
\begin{equation}
  \label{eq:dimensionless_variables}
  p=p_F+\frac{\pi T}{v_F}\,u,
  \quad
  \alpha=\frac{\sqrt3\pi T}{8\mu}(w+w'),
  \quad
  \alpha'=\frac{\sqrt3\pi T}{8\mu}(w-w')
\end{equation}
and denoting $\phi_p=\Phi(u)$.  Substitution of
Eq.~(\ref{eq:dimensionless_variables}) into
Eq.~(\ref{eq:W_operator_linearization}) yields $\widehat W
\phi_p=(3/32\pi^3\tau_a)\widehat\Omega \Phi(u)$, where
\begin{equation}
  \label{eq:Omega_operator}
  \widehat \Omega\Phi(u)=\!\iint \!dw\, dw'
  \frac{w^2+w'^2}
  {8w^2w'^2}
  G(u+w+w')G(u+w)G(u+w')
     \bigg(
     \frac{\Phi(u)}{G(u)}+\frac{\Phi(u+w+w')}{G(u+w+w')}
     -\frac{\Phi(u+w)}{G(u+w)}-\frac{\Phi(u+w')}{G(u+w')}
     \bigg),
\end{equation}
where the integrals extend from $-\infty$ to $+\infty$ and
\begin{equation}
  \label{eq:G}
  G(u)=\frac{1}{\cosh(\pi u/2)}.
\end{equation}
As a result the eigenvalue problem (\ref{eq:eigenvalue_problem})
takes the dimensionless form
\begin{equation}
  \label{eq:eigenfunctions}
  \widehat \Omega \Phi_l(u)=\omega_l\Phi_l(u),
\end{equation}
with the eigenvalues $\omega_l$ determining the relaxation rates
\begin{equation}
  \label{eq:relaxation_rates_relation}
  \frac{1}{\tau_l}=\frac{3\omega_l}{32\pi^3\tau_a}.
\end{equation}

Next, we use the symmetry $w\leftrightarrow w'$ to replace
$(w^2+w'^2)/w^2w'^2 \to 2/w^2$ in Eq.~(\ref{eq:Omega_operator}) and
perform the Fourier transform
\begin{equation}
  \label{eq:Fourier}
  \Phi(u)=\frac{1}{\sqrt{2\pi}}\int \varphi(x)e^{-iux}dx,
  \qquad
  \varphi(x)=\frac{1}{\sqrt{2\pi}}\int \Phi(u)e^{iux}du.
\end{equation}
Then  the eigenvalue problem (\ref{eq:eigenfunctions}) transforms to
\begin{equation}
  \label{eq:eigenvalue_problem_integral}
  \int_{-\infty}^{+\infty} K(x,x')\varphi_l(x')dx'=\omega_l\,\varphi_l(x),
\end{equation}
where the kernel is given by
\begin{equation}
  \label{eq:kernel-2}
  K(x,x')=\iint 
  \frac{du\,dw}
  {4\pi}
  \left[
    \frac{e^{iu(x-x')}}{w\sinh\frac{\pi w}{2}}
    \left(
       \frac{G(u+w)}{G(u)}-e^{-iwx'}
     \right)
     -2ie^{iux}\frac{\sin(wx')}{w^2\cosh x'}G(u+w)
  \right].
\end{equation}
Let us now split the kernel into three contributions:
\begin{equation}
  \label{eq:kernel}
  K(x,x')=K_1(x-x')+K_2(x,x')+K_3(x,x'),
\end{equation}
where
\begin{eqnarray}
  \label{eq:K1}
  K_1(x-x')&=&\int 
  \frac{du\,dw}
  {4\pi}
    \frac{e^{iu(x-x')}}{w\sinh\frac{\pi w}{2}}
    \left(
       \frac{G(u+w)}{2G(u)}+\frac{G(u-w)}{2G(u)}-1
     \right),
\\[1ex]
  \label{eq:K2}
  K_2(x,x')&=&\int 
  \frac{du\,dw}
  {4\pi}e^{iu(x-x')}
    \frac{1-\cos(wx')}{w\sinh\frac{\pi w}{2}}=\ln(\cosh x)\delta(x-x'),
\\[1ex]
  \label{eq:K3}
  K_3(x,x')&=&
  -\frac{i}{\cosh x'}
  \int \frac{du\,dw}
  {4\pi}e^{iux}
  \frac{\sin(wx')}{w^2}[G(u+w)-G(u-w)]
  =-\frac{|x+x'|-|x-x'|}{2\cosh x \cosh x'}.
\end{eqnarray}
Evaluation of the first kernel is somewhat nontrivial.  The result can
be presented in the form
\begin{equation}
  \label{eq:K1_result}
  \int_{-\infty}^\infty K_1(x-x')\varphi(x')dx'
  =\frac{1}{2}\int_{-\infty}^\infty
  \ln\left(2\tanh\frac{|x-x'|}{2}\right){\rm sgn\,}(x'-x)
  \frac{d\varphi}{dx'}dx'.
\end{equation}
In the following discussion we will only use the fact that $K_1$ is a
function of the difference $x-x'$; the explicit form
\eqref{eq:K1_result} will not be used.  The evaluation of the
integrals \eqref{eq:K2} and \eqref{eq:K3} is straightforward.

Note, that $\Phi_0(u)=G(u)$ is an obvious solution of the eigenvalue
problem \eqref{eq:eigenfunctions} with the eigenvalue $\omega_0=0$,
see Eq.~(\ref{eq:Omega_operator}).  Therefore, its Fourier transform
$\phi_0(x)=\sqrt{2/\pi}\,g(x)$, where
\begin{equation}
  \label{eq:g}
  g(x)=\frac{1}{\cosh x},
\end{equation}
must solve the eigenvalue problem
\eqref{eq:eigenvalue_problem_integral} with the same eigenvalue
$\omega_0=0$.  Noticing that $K_3(x,x')$ is odd in $x$ and $x'$, we
conclude that a condition
\begin{equation}
  \label{eq:K1_condition}
  \int_{-\infty}^\infty K_1(x-x')g(x')dx'=g(x)\ln g(x)
\end{equation}
must be satisfied.

We now establish some general properties of the derivatives of $g(x)$.
Noticing that
\begin{equation}
  \label{eq:g_properties}
  g'(x)=-g(x)\tanh x,
  \quad
  (\tanh x)'=1-\tanh^2 x,
\end{equation}
it is straightforward to show that the $l$-th derivative $g^{(l)}(x)$
is given by $g(x)$ multiplied by the polynomial of $\tanh x$ of $l$-th
power.  Let us then consider the most general function of this form:
\begin{equation}
  \label{eq:class}
  a_l^{} g^{(l)}(x)+a_{l-1}^{} g^{(l-1)}(x)+\ldots+a_0^{}g(x)
  =g(x)\big(b_l^{}\tanh^l x+b_{l-1}^{}\tanh^{l-1} x+\ldots+b_0^{}\big).
\end{equation}
We assume here that the leading coefficients $a_l^{}$ and $b_l^{}$ do not
vanish.
The two forms of the expression \eqref{eq:class} are equivalent; each
set of coefficients $(a_0^{}, a_1^{}, \ldots, a_l^{})$ uniquely
defines the set $(b_0^{}, b_1^{}, \ldots, b_l^{})$ and \emph{vice
  versa}.

We now show that when the integral operator with the kernel $K(x,x')$
is applied to a function of the form \eqref{eq:class}, the resulting
function also has form \eqref{eq:class}, with the same $l$.  By
applying the sum of $K_1$ and $K_2$ to the $l$-th derivative of $g(x)$
and using Eq.~\eqref{eq:K1_condition}, we find
\begin{eqnarray}
  \hspace{-2.5em}
  \int[K_1(x-x')+K_2(x,x')]g^{(l)}(x')dx'&=&
  \frac{d^l}{dx^l}\int K_1(x-x')g(x')dx'+g^{(l)}(x)\ln \cosh x
\nonumber\\
  &=&  \frac{d^l}{dx^l}[g(x)\ln g(x)]-g^{(l)}(x)\ln g(x)
\nonumber\\
  &=&\sum_{j=1}^l\frac{l!}{j!(l-j)!}g^{(l-j)}(x) \frac{d^j}{dx^j}\ln g(x).
\label{eq:binomial}
\end{eqnarray}
As we saw earlier $g^{(l-j)}(x)$ is given by $g(x)$ multiplied by a
polynomial of $\tanh x$ of power $l-j$.  Taking into consideration
Eq.~(\ref{eq:g_properties}) and noticing that $[\ln g(x)]'=-\tanh x$,
it is easy to see that $d^j\ln g(x)/dx^j$ is a polynomial of
$\tanh x$ of power $j$.  Thus each term in the last line of
Eq.~\eqref{eq:binomial} is $g(x)$ multiplied by a polynomial of
$\tanh x$ of power $l$, and therefore the right-hand side of
Eq.~\eqref{eq:binomial} has the form \eqref{eq:class}.  Addition of
the terms with lower-order derivatives, which are also present the
left-hand side of Eq.~\eqref{eq:class}, does not change the general
form of the result.  Thus, the application of the integral operator
with the kernel $K_1 +K_2$ to a function of form \eqref{eq:class}
gives a function of the same form.

Because $K_3(x,x')$ is odd in $x$ and $x'$, the corresponding integral
operator gives zero when applied to $g(x)\tanh^{2m}x$ with
$m=0,1,2,\ldots$.  Let us now apply this operator to
$g(x)\tanh^{2m+1}x$,
\begin{eqnarray}
  \label{eq:K3_calculation}
  \int K_3(x,x')g(x')\tanh^{2m+1}x'dx'
  &=&-\int \frac{|x+x'|-|x-x'|}{2\cosh x}
  \tanh^{2m+1}x'\frac{dx'}{\cosh^2x'}
\nonumber\\
  &=&\frac{g(x)}{4(m+1)}
  \int(\tanh^{2m+2}x'-1)
  [{\rm sgn}(x+x')-{\rm sgn}(x'-x)]dx'
\nonumber\\
  &=&\frac{g(x)}{2(m+1)}
  \int_{-x}^x(\tanh^{2m+2}x'-1)dx'
  =-\frac{g(x)}{m+1}\sum_{j=0}^m\frac{\tanh^{2j+1}x}{2j+1}.
\end{eqnarray}
\end{widetext}
Thus the integral operator with the kernel $K_3$ applied to a function
of the form \eqref{eq:class} results in a function of the same form.

We have therefore demonstrated that the action of the integral
operator with the kernel (\ref{eq:kernel}) on any function of the form
\eqref{eq:class} results in a function of the same form.  This enables
us to find the eigenfunctions of the integral operator in
Eq.~(\ref{eq:eigenvalue_problem_integral}).  We first notice that a
function of the form \eqref{eq:class} is fully described by $l+1$
coefficients $b_0^{}$, $b_1^{}$, \ldots, $b_l^{}$. Such functions form
an $(l+1)$-dimensional subspace, and our operator in this subspace is
a symmetric matrix of size $(l+1)\times(l+1)$. It has $l+1$
eigenfunctions that are orthogonal to each other and have the form
$g(x)p_l(\tanh x)$, where $p_l$ is a polynomial of power $l$.  When
$l$ is increased by 1, a new eigenfunction $g(x)p_{l+1}(\tanh x)$
appears.  Thus all the solutions have polynomials $p_l$ of different
powers.  The orthogonality condition
\begin{eqnarray}
  \label{eq:orthogonality_condition}
  &&\int_{-\infty}^{+\infty} g(x)p_l^{}(\tanh x)\,g(x)p_{l'}^{}(\tanh x)dx
  \nonumber\\
  &&\hspace{3em}=\int_{-1}^1 p_l^{}(y)p_{l'}^{}(y)dy=\delta_{l,l'}
\end{eqnarray}
indicates that $p_l^{}(y)$ are proportional to the Legendre
polynomials $P_l^{}(y)$.  The normalized eigenfunctions are
\begin{equation}
  \label{eq:eigenfunction_result}
  \varphi_l(x)=\sqrt{l+\frac12}\,\frac{P_l^{}(\tanh x)}{\cosh x}.
\end{equation}
Normalized eigenfunctions $\Phi_l(u)$ of the operator
(\ref{eq:Omega_operator}) are obtained by performing the inverse
Fourier transform (\ref{eq:Fourier}) of the above expression
\begin{equation}
  \label{eq:Phi_l}
  \Phi_l(u)=i^l\frac{\sqrt{2l+1}}{2\sqrt{\pi}}
  \int_{-\infty}^{+\infty}e^{-iux}
  \frac{P_l^{}(\tanh x)}{\cosh x}dx,
\end{equation}
where the additional factor $i^l$ ensures that $\Phi_l(u)$ is real for
all $l$.  Taking into account the definition of $u$ in
Eq.~(\ref{eq:dimensionless_variables}) and omitting the normalization
constant one obtains our result (\ref{eq:eigenfunctions-right}).

To find the eigenvalues $\omega_l$, we consider separately the cases
of even and odd $l$.  Because $K_3(x,x')$ is odd in $x$ and $x'$, it
does not affect the eigenvalues for even $l$, when the eigenfunction
(\ref{eq:eigenfunction_result}) is even in $x$.  The combined effect
of $K_1$ and $K_2$ on the eigenfunction $\varphi_l$ can be obtained
from Eq.~\eqref{eq:binomial}.  Its right-hand side is a linear
combination of $\varphi_j(x)$ with $j\leq l$.  Note that the term
$g(x)\tanh^lx$ appears only in $\varphi_l(x)$.  Thus the coefficient
of this term in the last line of Eq.~(\ref{eq:binomial}) is given by
that in the expansion of $g^{(l)}$ in the left-hand side times
$\omega_l$.  Using the relations (\ref{eq:g_properties}), one easily
finds
\begin{eqnarray*}
  \label{eq:leading-orders}
  \frac{g^{(l)}(x)}{g(x)}&=&(-1)^l\, l!\,\tanh^lx+\ldots,
  \\[0ex]
  \frac{d^j}{dx^j}\ln g(x)&=&(-1)^j (j-1)!\tanh^jx+\ldots,
\end{eqnarray*}
where the omitted terms have the form $c_j\tanh^j x$ with $j<l$.
Applying these results to Eq.~\eqref{eq:binomial}, we obtain
\begin{eqnarray*}
  &&\omega_l (-1)^l\, l!\, g(x)\tanh^lx
  =\sum_{j=1}^l\frac{l!}{j!(l-j)!}
\\
  &&\times(-1)^{(l-j)} (l-j)!\,g(x)\tanh^{l-j}x
  \,(-1)^j (j-1)!\tanh^jx.
\end{eqnarray*}
This immediately yields
\begin{equation}
  \label{eq:mu_even}
  \omega_l=\sum_{j=1}^l\frac{1}{j}
\end{equation}
for even $l$.

For odd $l=2m+1$ there is an additional contribution due to the $K_3$
part of the kernel.  It is given by the coefficient of $g(x)\tanh^{2m+1}x$
in the right-hand side of Eq.~\eqref{eq:K3_calculation}, i.e.,
$\delta \omega_{2m+1}=-[(m+1)(2m+1)]^{-1}$. Thus, for odd $l$ the
eigenvalue is
\begin{equation}
  \label{eq:mu_odd}
  \omega_l=\sum_{j=1}^l\frac{1}{j}-\frac{2}{l(l+1)}.
\end{equation}
Equations (\ref{eq:mu_even}) and (\ref{eq:mu_odd}) in combination with
Eq.~(\ref{eq:relaxation_rates_relation}) give the result
(\ref{eq:relaxation_rates}).

\section{Relaxation rates and modes in the spinless Fermi gas}
\label{sec:spinless}

Relaxation of the one-dimensional spinless Fermi gas was studied in
Ref.~\cite{matveev_thermal_2019}.  In the case of short-range
interaction the relaxation rates are given by
\begin{equation}
  \label{eq:relaxation_rates_spinless}
  \frac{1}{\tau_l}=\frac{2\pi^3\Lambda^2T^7}{\hbar^5 v_F^8}\lambda_l.
\end{equation}
Here the parameter $\Lambda$ is quadratic in $V(q)$ but different
from our earlier expression (\ref{eq:Lambda}); it is given by Eq.~(80)
of Ref.~\cite{matveev_thermal_2019}.  The parameters $\lambda_l$ are
obtained by solving the eigenvalue problem
\begin{equation}
  \label{eq:eigenvalue_problem_spinless}
  \widehat M \Phi_l(\xi)=\lambda_l \Phi_l(\xi).
\end{equation}
Here the operator $\widehat M$ is defined by
\begin{equation}
  \label{eq:K_alternative}
  \widehat M \Phi(\xi) = A(\xi) \Phi(\xi) 
           + \int d\xi'[B_1(\xi,\xi')+B_2(\xi,\xi')] \Phi(\xi'),
\end{equation}
where the integration is from $-\infty$ to $+\infty$ and
\begin{eqnarray}
  \label{eq:A}
  A(\xi)&=&\frac{(1+4\xi^2)(9+4\xi^2)(5+44\xi^2)}{5760},
\\[1ex]
  \label{eq:B1}
  B_1(\xi,\xi')&=&\frac{1}{6}(\xi-\xi')^2
             \frac{(\xi+\xi')[1+(\xi+\xi')^2]}{\sinh(\pi(\xi+\xi'))},
\\[1ex]
  \label{eq:B2}
  B_2(\xi,\xi')&=&-\frac{\xi-\xi'}{240\sinh(\pi(\xi-\xi'))}
\nonumber\\ &&\times
           \big(7+120\xi\xi'+128\xi^4-752\xi^3\xi'
\nonumber\\
           &&+1488\xi^2{\xi'}^2-752\xi{\xi'}^3
           +128{\xi'}^4\big).
\end{eqnarray}
Similarly to the relaxation problem (\ref{eq:eigenvalue_problem}) with
$\widehat W$ defined by Eq.~(\ref{eq:W_operator_linearization}), only
the right-moving particles are accounted for by the operator
$\widehat M$.  The relaxation of the left-moving particles can be
obtained by using the inversion symmetry of the system.

\begin{figure}[tb]
\includegraphics[width=.48\textwidth]{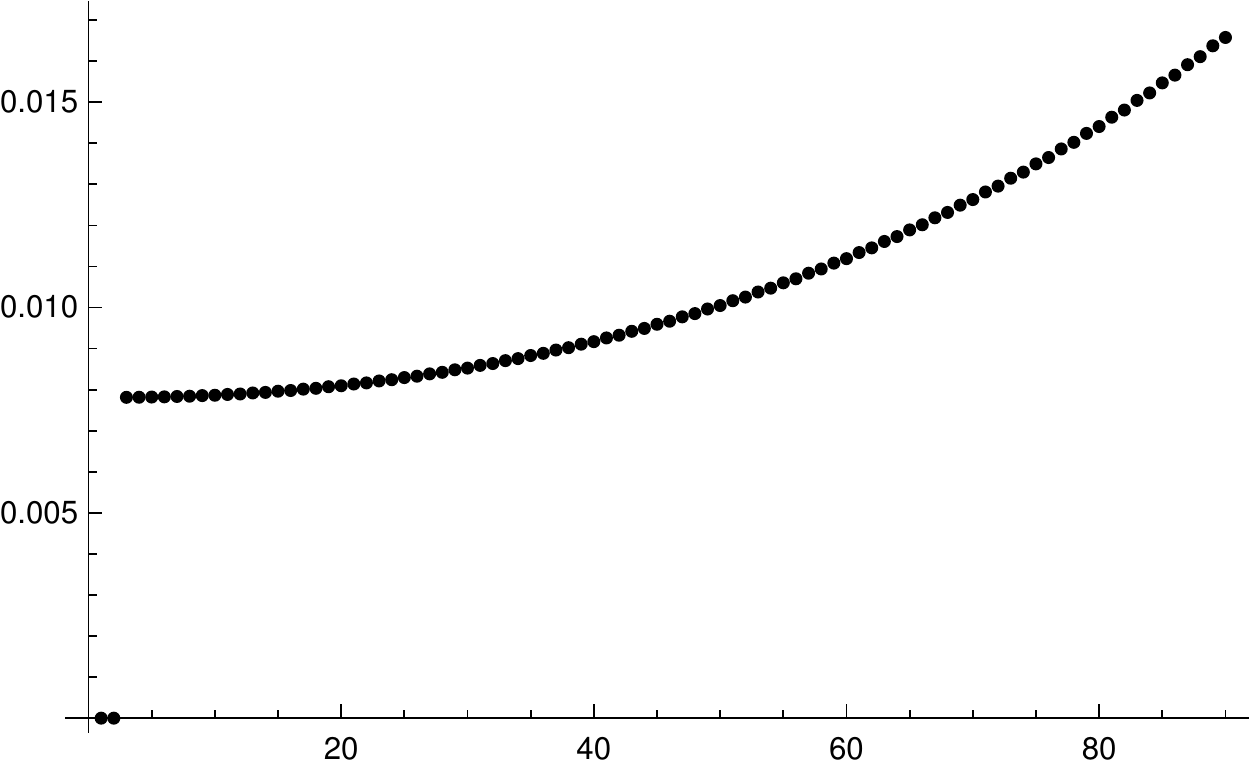}
\caption{Eigenvalues $\lambda_l$ of the operator $\widehat M$ obtained
  by numerical solution of the integral equation
  (\ref{eq:eigenvalue_problem_spinless}).}
\label{fig:spinless_eigenvalues}
\end{figure}

The integral equation (\ref{eq:eigenvalue_problem_spinless}) can be
solved numerically by replacing the infinite limits of integration
with finite but large ones and discretizing the function $\Phi(\xi)$.
The resulting spectrum of eigenvalues is shown in
Fig.~\ref{fig:spinless_eigenvalues}.  The two lowest eigenvalues
vanish.  A gap $\Delta=A(0)=1/128$ separates $\lambda_0=\lambda_1=0$
and $\lambda_2\approx 0.00781$.  The dense set of eigenvalues above
the gap represents a continuous spectrum of relaxation rates and
extends to $+\infty$.

\begin{figure}[b]
\includegraphics[width=.48\textwidth]{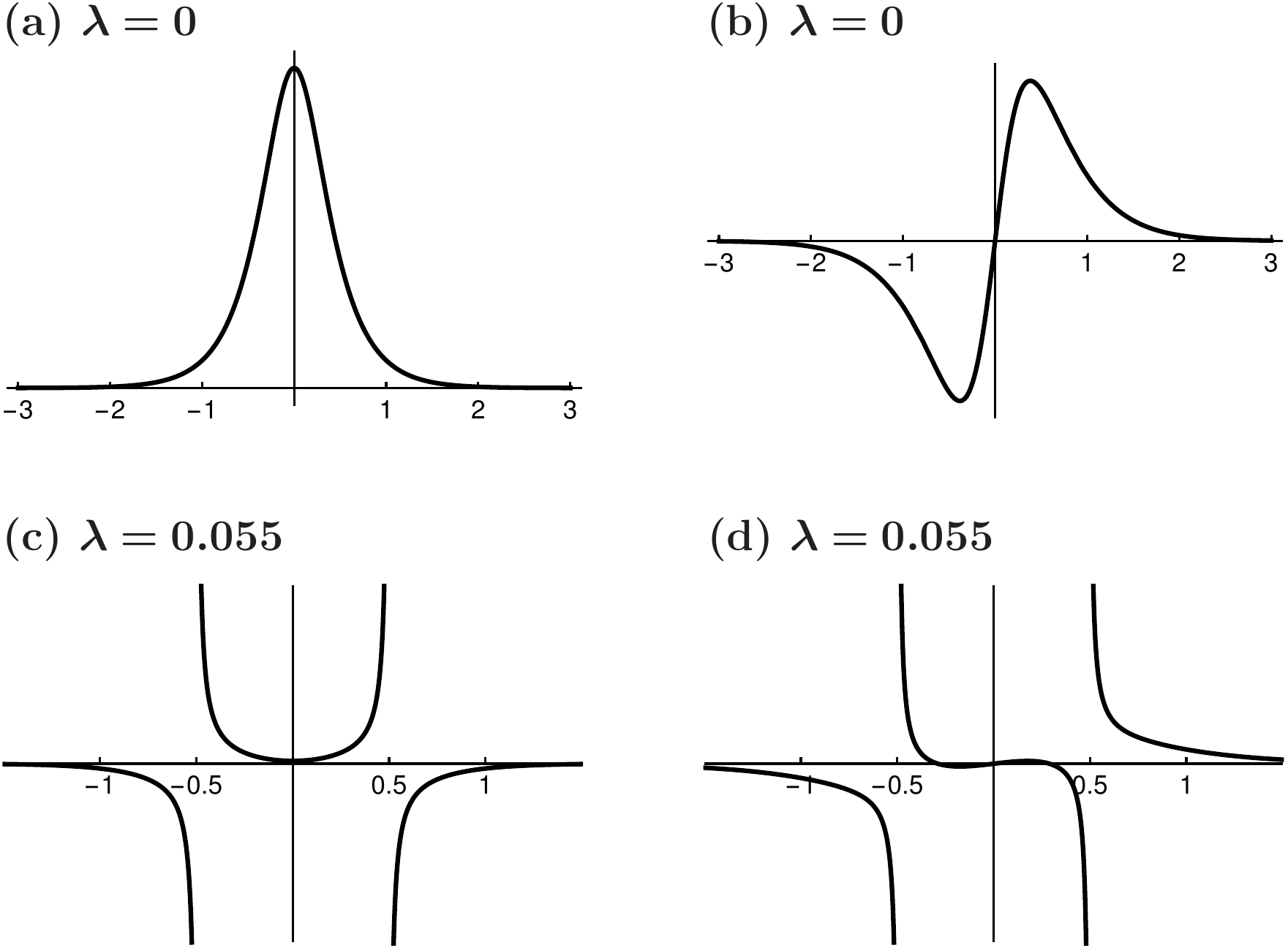}
\caption{Numerically obtained eigenfunctions $\Phi_l(\xi)$ of the
  operator $\widehat M$. (a), (b) The two modes with zero eigenvalues.
  (c), (d) The even and odd modes with nonzero eigenvalues.  The
  position of the singularity is $\xi_0=0.5$.}
\label{fig:modes}
\end{figure}

The two modes with zero eigenvalues are plotted in
Fig.~\ref{fig:modes}(a) and (b).  Up to a numerical prefactor they are
given by
\begin{equation}
  \label{eq:zero-modes-spinless}
  \Phi_0(\xi)=\frac{1}{\cosh(\pi \xi)},
  \quad
  \Phi_1(\xi)=\frac{\xi}{\cosh(\pi \xi)}.
\end{equation}
These two modes account for the conservation of the number of
particles and energy and are fully analogous to the modes
(\ref{eq:zero_modes}) for fermions with spin.

The modes corresponding to nonvanishing eigenvalues are qualitatively
different.  Each mode is either an even or an odd function of $\xi$
and has two singularities $(\xi\pm\xi_0)^{-1}$ for some value of
$\xi_0$.  The corresponding eigenvalue is related to the positions
$\pm\xi_0$ of the singularities by $\lambda=A(\xi_0)$.  Pairs of even
and odd modes are present for all real $\xi_0$.  Typical modes with
$\xi_0= 0.5$ are shown in Fig.~\ref{fig:modes}(c) and (d).

\bibliography{relaxation-spins}

\begin{thebibliography}{22}
\bibitem [{thr()}]{threefootnote}%
  \BibitemOpen
  \href@noop {} {}\bibinfo {note} {Polynomials $B_l(u)$ are related to the
  Bateman polynomials $F_l(u)$ via $B_l(u)=i^l F_l(iu)$}\BibitemShut {NoStop}%
\bibitem [{two()}]{twofootnote}%
  \BibitemOpen
  \href@noop {} {}\bibinfo {note} {We have multiplied the result (41) of
  Ref.~\cite{matveev_thermal_2019} by 2 to account for the spins of the
  fermions.}\BibitemShut {Stop}%
\bibitem [{fou()}]{fourfootnote}%
  \BibitemOpen
  \href@noop {} {}\bibinfo {note} {For spinless systems, it was recently
  predicted \cite{samanta_thermal_2019} that well below the quasiparticle
  relaxation rate the frequency-independent result $\kappa_{\rm ex}$ for the
  thermal conductivity of the system is replaced by
  $\kappa\propto\omega^{-1/3}$. We leave the study of this effect for systems
  of fermions with spin for future work.}\BibitemShut {Stop}%
\end{thebibliography}

\end{document}